# Snow water equivalent forecast maps seven days ahead from weather and remote sensing

Ziheng Sun[1*] and Hui Fang[2]

[1] Center for Spatial Information Science and Systems, George Mason University, Fairfax, VA 22030, USA

[2] Center for Satellite and Earth Science Research, George Mason University, Fairfax, VA, USA

*Corresponding author. Email: zsun@gmu.edu

**Abstract**

Operational snow water equivalent (SWE) maps guide water management, yet grid cells without ground stations lack the private pillow history that station models exploit. This study trains a soft constrained mix adjacency graph convolutional network on 49 SNOw TELemetry pillows. On each forecast date the network takes one feature vector of length 52 per pillow from information known by the end of that day, including up to 21 days of pillow and weather history, neighbour SWE, terrain, snow index, and forecast date priors, then emits one pillow forecast and one regional map for SWE on the target date seven days ahead, the calendar day the forecast is for. Fitting prepares the model for those cells by hiding pillow SWE history, blanking nearby pillow groups, scoring temporary locations without ground stations, and raising loss weight at isolated pillows. Snow Data Assimilation System (SNODAS) and University of Arizona (UA) analyses never enter as inputs. Sixteen pillows and four winters are withheld from fitting. On 11006 held pillow days RMSE is 0.936 inches versus the seasonal change prior at 1.181 inches, a relative RMSE gain of 20.7 percent. Maps compare the forecast with SNODAS and UA for mountain versus desert structure.

## 1 Introduction

Snow water equivalent (SWE) is the liquid depth that a snowpack would yield if melted, and it controls much of the warm season water supply in the western United States (Bales et al., 2006; Sturm et al., 2017). Contiguous daily SWE maps from the Snow Data Assimilation System (SNODAS) of the National Operational Hydrologic Remote Sensing Center (2004) and from the University of Arizona (UA) daily SWE dataset (Broxton et al., 2016; Zeng et al., 2018) are the operational analysis and reconstructed fields most often compared with pillows, surveys, and lidar (Clow et al., 2012; Painter et al., 2016; Dawson et al., 2018). Those products describe SWE on a calendar day after assimilation or reconstruction; they are not numerical forecasts of SWE several days ahead. The SNOw TELemetry (SNOTEL) network supplies the pillow record that anchors those comparisons (Serreze et al., 1999). Station learning has advanced short lead pillow forecasts from weather and related covariates (Kratzert et al., 2018; Song et al., 2024), and graph convolutional networks of the Kipf and Welling (2017) subtype can share information among irregular nodes (Wu et al., 2021). Soft constraints place process knowledge in the training loss rather than as a hard projection (Raissi et al., 2019; Karniadakis et al., 2021).

The gap addressed here is a direct regional forecast map, not a product correction, and not a pillow only network that is later pasted onto a grid. A network that ingests SNODAS or UA on the forecast date is a residual corrector of those analyses. A network that only minimises error at dense SNOTEL pillows can memorise each pillow's private SWE history and neighbour SWE history that map cells never have. The experiment instead forecasts pillow SWE from a forecast date feature vector that summarises up to 21 days of pillow and weather history ending on day $t$, together with neighbour SWE, optical snow index, and terrain, never from SNODAS or UA, adds fitting steps that force the network to behave like a grid cell without a ground station, and writes the map surface with a terrain densification workflow that is separate from the pillow MixGCN forward pass. SNODAS and UA appear only as map comparison rows. Pillows withheld from fitting and evaluated only after the model is fixed stand in for that setting: a new gauge must be served from neighbours without a fitting archive.

The aim is therefore to train a soft constrained mix adjacency graph network on SNOTEL pillows with weather and remote sensing inputs, emit one regional forecast map of SWE on target date $t + 7$ that remains accurate away from the dense network, and report relative RMSE gain against the best classical forecast date baseline on spatially withheld pillows. Classical references include persistence, meaning initial SWE on the forecast date held fixed to the target date, and the seasonal change prior already named above. SNODAS and UA remain daily analysis fields for geographic comparison in the map figures: the scientific claim is a seven day forward forecast map that approaches the mountain versus desert structure of those analyses without using them as inputs or as relative RMSE gain baselines.

## 2 Related work

### *2.1 Operational SWE maps and pillow evaluation*

Because the claim is a forecast map rather than a product correction, the nearest prior literature is how operational SWE analyses are built and how they are compared with pillows. SNODAS assimilates station and remotely sensed snow information into a contiguous analysis of the calendar day (National Operational Hydrologic Remote Sensing Center, 2004). The UA daily SWE product reconstructs snowfall and accumulation into spatial maps that have been compared with snow cover extent and pillow records (Broxton et al., 2016; Dawson et al., 2018; Zeng et al., 2018). Independent evaluations against Colorado Rockies surveys and airborne snow observatory flights show where analyses agree with and diverge from pillows and lidar (Clow et al., 2012; Painter et al., 2016). Neither product is a multi day forward SWE forecast. The present paper therefore places SNODAS and UA analyses dated to the forecast date $t$ beside the proposed map of target date $t + 7$ for geographic quality comparison, never as forecasting competitors and never as network inputs.

### *2.2 Station SWE forecasting*

Product comparisons alone do not decide whether a learned model can forecast pillows that never entered fitting, so station SWE forecasting is the next strand. Recurrent networks and related machine learning models now forecast daily SWE and runoff at gauged sites when a fitting archive exists (Kratzert et al., 2018; Song et al., 2024). Persistence and explicit hydrological benchmarks remain the standard checks against overstated forecast accuracy (Murphy, 1993; Pappenberger et al., 2015). Most station studies evaluate pillows that supplied training rows. The evaluation here asks whether neighbours and weather alone can serve pillows that never entered fitting, which is the pillow check used for the map claim.

### *2.3 Graph networks and soft physics constraints*

Serving held pillows from neighbours requires a graph operator, and keeping pack physics in training requires a soft constraint, which is the third strand. Graph convolutional networks of the Kipf and Welling subtype update node states through a normalised adjacency operator and are a standard subtype within the broader graph neural network literature (Kipf and Welling, 2017; Wu et al., 2021). Soft constraints add a physics residual to the training loss rather than projecting outputs after training (Raissi et al., 2019; Karniadakis et al., 2021). The network used here is a residual mix adjacency graph convolutional network with a soft constraint weight selected on validation winters and with fitting steps that discourage memorising dense SNOTEL archives, while the map surface uses a separate terrain densification workflow for grid cells without ground stations.

## 3 Methods

### *3.1 Forecast maps and spatially withheld pillows*

With that literature in view, the experiment forecasts SWE on target date $t + 7$ while withholding entire pillows from fitting. The forecast date is the calendar day $t$ when the forecast is produced. The target date is day $t + 7$, when the forecast is compared with the pillow observation. One forward pass on day $t$ yields one pillow forecast $\hat{S}_{i,t+7}$ and, after terrain densification, one regional map of that same target date field. The target is pillow SWE $S_{i,t+7}$ in inches at pillow $i$ on the target date. Initial SWE is the pillow observation $S_{i,t}$ on the forecast date. Forecast dates run from November through April. Entire pillows are withheld from fitting so that evaluated sites stand in for map cells that never supplied a private fitting archive. Sixteen of 65 pillows are held pillows. They contribute no rows to fitting, early stopping, selection of the soft constraint weight, or the seasonal change prior. The other 49 are fit pillows. Held winters are 2020, 2021, 2022, and 2024. Fitting winters are 2012 through 2017 and validation winters are 2018 and 2019. The held sample has 11006 pillow days at held pillows.

The held set is not redesigned for this map paper. The sixteen station identifiers were drawn once at random with a fixed seed in the sister pillow network study on the same 65 site table, then frozen here so that later fitting choices cannot retune which gauges enter the relative RMSE gain comparison. About one quarter of the network is held (16 of 65), which is large enough for multi year pillow day counts and still leaves a dense enough fit graph for neighbour message passing. After the draw, the held pillows happen to span Arizona (six), New Mexico (six), southern Colorado (three), and Nevada (one), and elevations from about 2320 m to 3676 m, so the evaluation is not confined to one county or one elev belt. Distance from each held pillow to the nearest fit pillow ranges from a few kilometres inside dense clusters to more than 100 km at sparse sites, which later supports the distance stratum check.

Temporal splits follow the same frozen calendar. Fitting winters (2012 through 2017) build weights and the seasonal change prior. Validation winters (2018 and 2019) select the soft constraint weight and stop training. Held winters (2020, 2021, 2022, and 2024) are used only after the checkpoint is fixed. Year 2023 is not in the held list in this package. That ordering keeps climate priors and early stopping blind to the held winters.

A forecast date baseline may use only information available at the end of the forecast date. Persistence copies initial SWE to the target date. The seasonal change prior, also called the bin prior, adds the fitting mean seven day SWE change for that five day calendar window to initial SWE. SNODAS and UA cells are never used as baselines for the relative RMSE gain threshold because they are daily analysis and reconstructed fields, not forward forecasts from day $t$ to day $t + 7$, and because they are not allowed as model inputs. In the map figures they serve only as geographic quality references for the proposed $t + 7$ forecast map.

### *3.2 Study area and data*

Those timing and holdout rules are applied inside one fixed southwestern United States box from about 32.6 to 37.6 degrees north and 116.2 to 104.8 degrees west. It contains 65 SNOTEL pillows in Arizona, New Mexico, southern Colorado, and the Spring Mountains of Nevada, at elevations from about 1787 m to 3676 m. Figure 2 places that box on a western United States locator and then expands the same extent on cloudless true colour so fit pillows, held pillows, and nested map windows can be read before any densified map is judged.

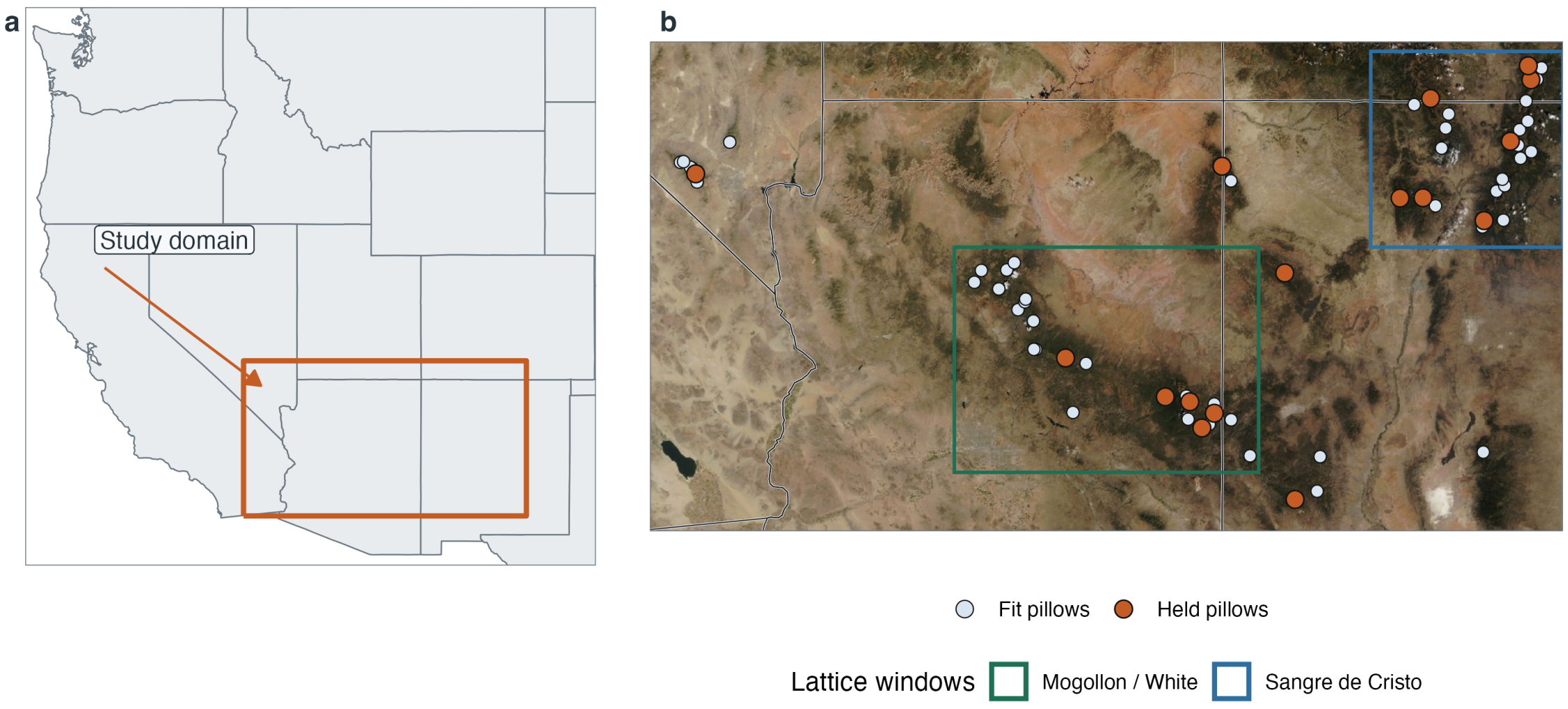

Figure 2. Western locator with the study domain box (a) and Southwest true colour with fit versus held pillows and nested map windows (b).

Part (a) outlines the southwestern box on a western United States locator; an arrow marks the domain used for every map product in this paper. The box mixes mountain snow that matters for spring water supply with desert and foothill floors where densified maps can be judged on whether snow stays on ridges rather than painting basins; the 65 pillows form a connected SNOTEL graph dense enough for neighbour message passing while still leaving room for the 16 pillow spatial holdout already named above; daily SNODAS and UA fields cover the box continuously for comparison in the map figures even though those fields never enter the model; and the domain matches the frozen pillow table and held draw of the sister station study so map claims and withheld pillow claims share one geography. Part (b) is cloudless MODIS true colour inside the same box. Light markers are the 49 fit pillows used for training, early stopping, and the seasonal change prior. Dark markers are the 16 held pillows reserved for evaluation after the model is fixed and never used in fitting. Blue and green rectangles mark the Sangre de Cristo and Mogollon Rim / White Mountains map windows nested inside the full Southwest extent. The full Southwest extent carries the regional forecast surface; the two subwindows zoom on mountain cores where pillow density and winter pack contrast are high. Pillows outside those subwindows still enter the held table; the subwindows are display cuts, not a second experiment.

Pillow SWE is the Natural Resources Conservation Service water equivalent field (swe_value) in inches, the native SNOTEL reporting unit (1 inch = 25.4 mm). An alternate column labelled SWE in the source table correlates only weakly with swe_value and is never used as the target. Weather type fields follow gridMET style inputs (Abatzoglou, 2013). Normalised difference snow index, terrain attributes, and pillow lag and neighbour fields complete the feature set. SNODAS and UA collocations are stored for comparison in the map figures only and are excluded from every feature column. True colour tiles for the map figures are from NASA Global Imagery Browse Services and are optical context only.

### *3.3 Feature construction and forecast date priors*

Inside that domain the proposed model is not a recurrent sequence net over daily frames. For each November through April forecast date $t$ and pillow $i$, it receives one fixed length feature vector of dimension $F = 52$ that is assembled only from quantities known at the end of day $t$. The target is pillow SWE on target date $t + 7$, written $S_{i,t+7} = \text{swe_value}_{i,t+7}$. The longest lookback used to build those features is 21 days ending on the forecast date: rolling means and sums use windows of 3, 7, 14, and 21 days on the series through day $t$, and the explicit SWE lag stack uses days $t - 1$ through $t - 14$ together with the forecast date value $S_{i,t}$. No observation from day $t + 1$ through $t + 7$ enters any feature.

The 52 columns fall into eight groups that match the deposited checkpoint feature list. Local pillow SWE state contributes twenty one columns: initial SWE $S_{i,t}$, fourteen lag values $S_{i,t-1}, \dots, S_{i,t-14}$, the one day and seven day differences $S_{i,t} - S_{i,t-1}$ and $S_{i,t} - S_{i,t-7}$, and rolling SWE means over 3, 7, 14, and 21 days. Neighbour SWE on the forecast date contributes one column $\text{nei}_{i,t}$ defined below. Forecast date priors contribute five columns: anomaly persistence, day of year climatology of the target date, the seasonal change prior, the temperature index prior, and the soft constraint blend $\Phi_{i,t}$. Weather history contributes eight columns: precipitation sums and mean air temperature over the same 3, 7, 14, and 21 day windows ending on $t$. Same day weather contributes seven gridMET style fields on day $t$: mean air temperature $T_{i,t}$ (Celsius), precipitation $P_{i,t}$, daily maximum and minimum air temperature, relative humidity extremes, and wind speed. Static terrain contributes five columns: elevation $z_i$, slope, aspect, northness, and eastness. Optical snow cover contributes $\mathrm{NDSI}_{i,t}$. Location and calendar contribute four columns: latitude, longitude, day of year of the forecast date, and day of year of the target date $t+7$ (a calendar quantity known when the forecast is made, not an observation of SWE after $t$). All SNODAS and UA columns are dropped before fitting, so none of the 52 features is an operational SWE analysis.

A neighbour SWE covariate is built from the same geographic edge weights later used for the MixGCN adjacency. With raw edge weight $w_{ij}$ based on distance and elev difference, and with $A^{\mathrm{nb}} = A - \mathrm{diag}(A)$, the neighbour mean on day $t$ is

$$\mathrm{nei}_{i,t} = \frac{\sum_{j\neq i} A^{\mathrm{nb}}_{ij}\, S_{j,t}}{\sum_{j\neq i} A^{\mathrm{nb}}_{ij}}.$$

Two forecast date priors are computed from fit pillows and fit winters only. The temperature index prior unrolls a simple pack update for seven steps with melt factor $m = 0.08$ inch °C$^{-1}$ day$^{-1}$. Starting from $s^{(0)} = S_{i,t}$,

$$s^{(k)} = \max\big\{0,\ s^{(k-1)} + P_{i,t}\,\mathbf{1}(T_{i,t} \leq 0) - m\max(0, T_{i,t})\,\mathbf{1}(T_{i,t} > 0)\big\}, \qquad k = 1, \ldots, 7,$$

and $\tilde{S}^{\mathrm{TI}}_{i,t+7} = s^{(7)}$. Forecast date weather is held fixed from day $t$ to the target date inside that unroll. The seasonal change prior (bin prior) uses five day calendar bins $b(t) = \lfloor (d(t) - 1)/5 \rfloor$ for day of year $d(t)$. Let $\bar{\Delta}_b$ be the mean of $S_{i,t+7} - S_{i,t}$ over fit pillows and fit years in bin $b$. Then

$$\tilde{S}^{\mathrm{bin}}_{i,t+7} = \max\{0,\ S_{i,t} + \bar{\Delta}_{b(t)}\}.$$

Held pillows never contribute to $\bar{\Delta}_b$. Day of year climatology $\tilde{S}^{\mathrm{clim}}_{i,t+7}$ is the fit sample mean of pillow SWE on the target date day of year, and anomaly persistence adds the forecast date anomaly relative to the forecast date climatology to that target date climatology. The soft constraint target blends the temperature index and seasonal change priors with fixed weights

$$\Phi_{i,t} = 0.55\, \tilde{S}^{\mathrm{TI}}_{i,t+7} + 0.45\, \tilde{S}^{\mathrm{bin}}_{i,t+7}.$$

Both priors, climatology, anomaly persistence, and $\Phi$ also appear among the $F = 52$ input columns, so the network sees them as covariates as well as through the soft loss. Features are standardised with means and standard deviations estimated on fit pillow rows only. Missing required columns drop the row.

*3.4 Station graph and MixGCN forward pass*

Those standardised features sit on a fixed station graph rather than on independent rows. Nodes are the $N = 65$ pillows. For each pillow the $K = 14$ nearest neighbours in longitude and latitude define candidate edges with

$$w_{ij} = \exp\left(-\frac{d_{ij}}{0.65}\right)\exp\left(-\frac{|z_i - z_j|}{450}\right),$$

where $d_{ij}$ is distance in degrees and $z$ is elevation in metres. The matrix is symmetrised by $\max(w_{ij}, w_{ji})$, a self loop of weight one is added to form $A$, and the Kipf and Welling normalised adjacency is $\tilde{A} = D^{-1/2} A D^{-1/2}$ with $D_{ii} = \sum_j A_{ij}$. The graph is fixed from coordinates and elevation only.

The proposed network is a residual MixGCN. With a learnable scalar $a$ initialised at 0.45 and $\alpha = \sigma(a)$, the mixed propagator is

$$\hat{A} = \alpha \tilde{A} + (1 - \alpha) I.$$

Standardised node features $X \in \mathbb{R}^{N \times 52}$ enter a three layer encoder (widths $52 \rightarrow 320 \rightarrow 320 \rightarrow 320$) with GELU and dropout 0.12, yielding $H^{(0)}$. Three residual graph layers then apply

$$H^{(l)} = \mathrm{LN}\left(\mathrm{Drop}\left(\mathrm{GELU}(\hat{A}\, H^{(l-1)} W^{(l)})\right)\right) + H^{(l-1)}, \qquad l = 1,2,3,$$

with $W^{(l)} \in \mathbb{R}^{320 \times 320}$ and layer normalisation. A head $320 \rightarrow 160 \rightarrow 1$ produces a change $\delta_i$. With learned scale $\gamma$ (initialised at one), the forecast is

$$\hat{S}_{i,t+7} = \max\{0,\ S_{i,t} + \gamma\, \delta_i\}.$$

The head therefore predicts a seven day change anchored on initial SWE, not an absolute level. On each day the full $N \times N$ graph is filled; pillows without a record that day carry a zero standardised feature row. Held pillow RMSE uses this station forward pass alone. The matched identity ablation replaces $\hat{A}$ by $I$ in every graph layer so neighbour mixing cannot enter through adjacency. The $\lambda = 0$ ablation keeps MixGCN but removes the soft physics term. Figure 1 gathers that forward pass with the training loss and the map densification path that still need to be defined.

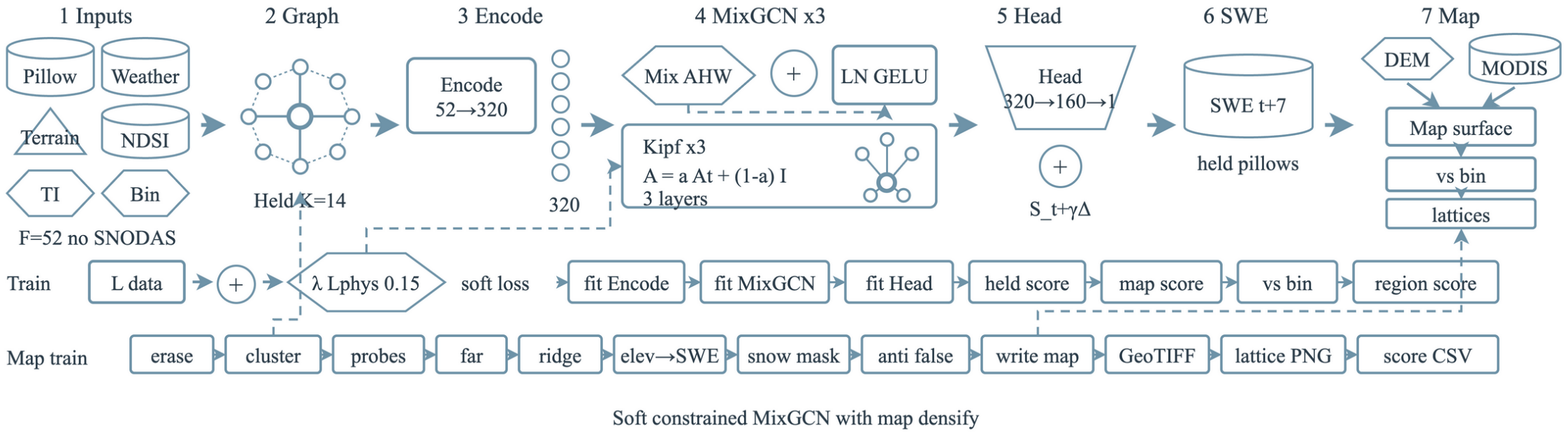

Figure 1. Forward pass from forecast date inputs through MixGCN layers to SWE on target date $t + 7$, with soft loss and terrain densify under the spine.

Figure 1 is read left to right in seven numbered stages, then two training rows under the spine. Stage 1 (Inputs) stacks the $F = 52$ feature vector described in the feature construction section: up to 21 days of pillow SWE and weather history ending on forecast date $t$, neighbour SWE on $t$, terrain, normalised difference snow index (NDSI), and the forecast date priors (temperature index TI, seasonal change Bin, climatology, anomaly persistence, and soft blend), with no SNODAS or UA columns and no sequence of daily frames after $t$. Stage 2 (Graph) places the held pillows on the fixed station graph with $K = 14$ neighbours so message passing can serve sites that never entered fitting. Stage 3 (Encode) maps the 52 features to a width 320 node state with GELU and dropout. Stage 4 (MixGCN x3) applies three residual mix adjacency graph layers with layer normalisation and GELU, using $\hat{A} = \alpha \tilde{A} + (1 - \alpha) I$ as drawn in the Kipf block. Stage 5 (Head) maps 320 to 160 to 1 and forms the residual forecast $\hat{S}_{i,t+7} = \max\{0, S_{i,t} + \gamma \delta_i\}$ shown on the canvas as $S_t + \gamma \Delta$. Stage 6 (SWE) is the target date water equivalent $\hat{S}_{i,t+7}$ at pillows withheld from fitting. Stage 7 (Map) densifies that same target date field: NASADEM and forecast date

MODIS drive terrain densification to a map surface for day $t + 7$ that is later compared with the seasonal change prior (vs bin) in the map figures; MixGCN station blobs are not pasted onto the grid.

The Train row under the spine joins the data loss $L_{\text{data}}$ with the soft physics term $\lambda L_{\text{phys}}$ at $\lambda = 0.15$, then fits Encode, MixGCN, and Head in one run, then scores withheld pillows and regional maps against the seasonal change prior. The Map train densify row states the fitting steps for grid cells without ground stations and the densify write path in order: hide private pillow SWE history, blank nearby fit pillow groups, score temporary locations without ground stations, raise loss weight at isolated fit pillows, score ridges, convert elevation prominence to SWE depth, apply the snow mask, remove false snow freckles, and write GeoTIFF, map panel PNG, and score CSV products. The soft constraint remains a training loss penalty, not a post training blend. The loss and densification equations that follow let each of those blocks be reimplemented.

*3.5 Soft constraint loss and training for grid cells without ground stations*

Because neighbour mixing alone can still overfit dense SNOTEL identity, the training objective must pull forecasts toward the physics prior $\Phi$ while also simulating map cells. For supervised fit nodes $\mathcal{M}$ on a day, with mask $m_i = 1$, let $\ell$ be the Huber smooth L1 map ($\ell(e) = e^2/2$ for $|e| < 1$ and $|e| - 1/2$ otherwise). The data term is an asymmetric weighted mean

$$L_{\text{data}} = \frac{\sum_{i\in\mathcal{M}} w_i \ \ell(\hat{S}_{i,t+7} - S_{i,t+7})}{|\mathcal{M}|},$$

$$w_i = \left(1 + 1.15\,\mathbf{1}(\hat{S}_{i,t+7} < S_{i,t+7})\right)\left(1 + 1.2\,\mathbf{1}(S_{i,t} \geq 10) + 1.8\,\mathbf{1}(|S_{i,t+7} - S_{i,t}| \geq 2)\right),$$

so underestimates are upweighted to 2.15, deep packs on the forecast date (at least 10 inches) receive a further factor 2.2, and large seven day changes (at least 2 inches) receive a further factor 2.8. The physics term is

$$L_{\text{phys}} = \frac{\sum_{i\in\mathcal{M}} \ell\,(\hat{S}_{i,t+7} - \Phi_{i,t})}{|\mathcal{M}|}, \qquad L = L_{\text{data}} + \lambda L_{\text{phys}}.$$

The weight $\lambda$ is chosen from $\{0, 0.05, 0.1, 0.15, 0.2\}$ by validation RMSE on fit pillows in 2018 and 2019. Selected $\lambda = 0.15$ (validation RMSE about 0.773 inch) is used for the proposed ensemble and matches the Train row of Figure 1.

Even with that soft pull, a checkpoint fit only on dense pillows can still fail on grid cells without ground stations, so fitting adds four mechanisms that match the short labels erase, cluster, probes, and far on Figure 1. Hiding private pillow SWE history, with probability $p_{\text{erase}} = 0.32$, blanks local SWE history columns (lags, differences, and rolling SWE) on a random subset of supervised fit pillows and replaces them by the neighbour field path used for map cells, while initial SWE is kept for the residual anchor. Blanking nearby fit pillow groups, with probability $p_{\text{cluster}} = 0.12$ per day, applies the same blanking to an entire geographic group of fit pillows so the network cannot lean on a crowded neighbourhood. Scoring temporary locations without ground stations attaches $n_p = 12$ lon lat points with no pillow inside the study box each training day. Weather, NDSI, and terrain at those points are inverse distance weighted from fit pillows,

$$x_c^{\text{probe}} = \frac{\sum_{j\in\text{fit}} e^{-(d_{cj}/\ell)^2}\, x_j}{\sum_{j\in\text{fit}} e^{-(d_{cj}/\ell)^2}},$$

with length scale $\ell = 0.55°$. Those temporary locations attach to the graph through edges to the six nearest fit stations. Only a soft physics loss toward $\Phi$ is applied there, with weight 0.10, so

$$L_{\text{map}} = L_{\text{data}} + \lambda L_{\text{phys}} + 0.10\, L_{\text{probe}}.$$

Raising the loss weight by 1.35 at fit pillows whose mean neighbour spacing exceeds the fitting median keeps isolated fit pillows from being ignored. Together these steps train for grid cells without ground stations rather than for memorising dense pillow archives.

### *3.6 Optimisation and ensemble scoring*

With the loss and those fitting steps for cells without ground stations fixed, Encode, MixGCN, and Head are fit jointly in one AdamW run rather than as three separate jobs. Learning rate is $7 \times 10^{-4}$, weight decay $2 \times 10^{-4}$, with cosine annealing and gradient norm clipping at 2.0. Each epoch shuffles fit days. Early stopping waits eight epochs without a validation improvement of $10^{-4}$ inch, with a budget of 40 epochs for the proposed network. At the selected $\lambda$, a pool of seeds is trained; the ten members with lowest validation RMSE are kept and averaged with weights proportional to the inverse validation RMSE. At test time each member is evaluated with twelve forward passes under dropout (test time averaging). No held tuned residual scale is applied after the ensemble. Matched identity and $\lambda = 0$ ensembles use the same recipe with smaller seed pools.

Once that ensemble is frozen, fit and held pillows remain on the graph for neighbour context, but only held pillows and held winters enter the primary RMSE table. Map score and vs bin then evaluate the densified surface against the seasonal change prior and against SNODAS analysis geography for diagnostics in the map figures only. Region score is the pooled held pillow RMSE inside the southwestern study domain that later appears in Results.

### *3.7 Map densification preprocessing and postprocessing*

Pillow RMSE from that station path is not yet a regional map, so surfaces for the map figures are written afterward by a separate terrain densification workflow on the native product grid (about $1/24°$ cells). MixGCN station blobs are not pasted onto the grid. SNODAS and UA never enter features or gates; SNODAS analysis tip statistics on the forecast date are used only as an optional soft tip depth reference after morphology is fixed. This is the ridge through write map sequence on the bottom row of Figure 1.

For each forecast date shown in the map figures, NASADEM elevation $z(c)$ is first placed on the product grid. DEM slope $\theta(c)$ (degrees) and local prominence $\pi(c) = z(c) - \bar{z}_{5\times5}(c)$ are computed by finite differences and a five cell neighbourhood mean. A ridge score

$$r(c) = \tilde{z}(c)\left(0.55\, \tilde{\pi}(c) + 0.45\, \tilde{\theta}(c)\right)$$

normalises elev, prominence, and slope into [0,1] bands. Forecast date MODIS NDSI is collocated where clear. Fit pillow weather is transferred to cells by inverse distance weighting, then air temperature is lapse corrected toward elevation,

$$T(c) = \beta\, T^{\text{nn}}(c) + (1-\beta)\, T^{\text{IDW}}(c), \qquad T^{\text{nn}}(c) = T_{j^*(c)} - 0.0065\left(z(c) - z_{j^*(c)}\right),$$

with $\beta$ increasing with elev (about 0.45 to 0.95) and $j^*(c)$ the nearest fit pillow. An elev to SWE curve $f(z)$ is fit on snowy fit pillows (bin upper percentiles, linear elev slope, and elev rank to pillow quantiles), with a soft snowline ramp so pack just above the snowline stays light.

Morphology and depth postprocessing then write the map. Flat mid elev shelves with low prominence, shallow SWE, and clear low NDSI are set to zero. Snow presence grows only inside a ridge corridor

$$\mathcal{C} = \{c: z(c) \geq z_{\text{line}} - 100,\ \pi(c) \geq 10,\ \theta(c) \geq \theta_{\min},\ T(c) < 6.5,\ r(c) \geq 0.10\}$$

union clear optical mid slope snow under cold lapse temperatures. Binary dilation of snow cores is allowed only inside $\mathcal{C}$ (about four to five steps; fewer on sparse packs). Depth follows $f(z)$ with ridge weighted texture, elev band calibration to pillow means and upper tails, and an inside mask taper: cells near the pack edge stay shallow while interior distance into the pack allows deeper cores. Soft rim buffers outside $\mathcal{C}$ are not applied. Sparse days (snow fraction below about 0.07 to 0.12) refill cold prominent ridges from $f(z)$; snowy days strip mid shelf false snow and reinforce tip cores. Anti false snow kills clear bare foothills, warm low elev cells, and low prominence mid elev shelves with weak NDSI, then removes isolated freckles. Mild elev guided smoothing acts only inside the snow mask. Soft tip restore scales the top elev tenth of snowy cells toward SNODAS analysis tip percentiles on the forecast date for geographic realism without using SNODAS as an input feature or as a forward forecast.

After morphology and depth are fixed, the field is exported as a GeoTIFF on the product grid, rendered as the manuscript map PNG panels used in Results, and summarised in a score CSV of map diagnostics (snowy cell correlation with the SNODAS analysis on the forecast date, false snow rate, salt and pepper rate, tip depth ratio). Those products support the Map surface, vs bin, and map figure blocks at the right of Figure 1; they do not retune the held pillow relative RMSE gain.

*3.8 Scores*

Held pillow accuracy and map diagnostics are finally summarised with the same metrics used in Results. Primary RMSE is in inches on held pillow days,

$$\text{RMSE} = \sqrt{\frac{1}{n}\sum_{k=1}^{n}(\hat{S}_k - S_k)^2}.$$

Relative RMSE gain against a baseline $b$ is $G = (\text{RMSE}_b - \text{RMSE}_p)/\text{RMSE}_b$. The relative RMSE gain threshold for this package is $G \geq 0.20$ against the best deposited classical forecast date baseline on the frozen held sample. Map surface diagnostics (snowy cell correlation with the SNODAS analysis on the forecast date, false snow rate, salt and pepper rate, tip depth ratio) are reported for geographic quality against daily analyses only and are not used to retune the held pillow relative RMSE gain comparison.

## 4 Results

*4.1 Spatial pattern of relative gain*

With the southwestern domain and the fit versus held pillows fixed in Figure 2, the first held check asks whether relative RMSE gain versus the seasonal change prior is regional or local. Figure 3 colours the 16 held pillows by that gain.

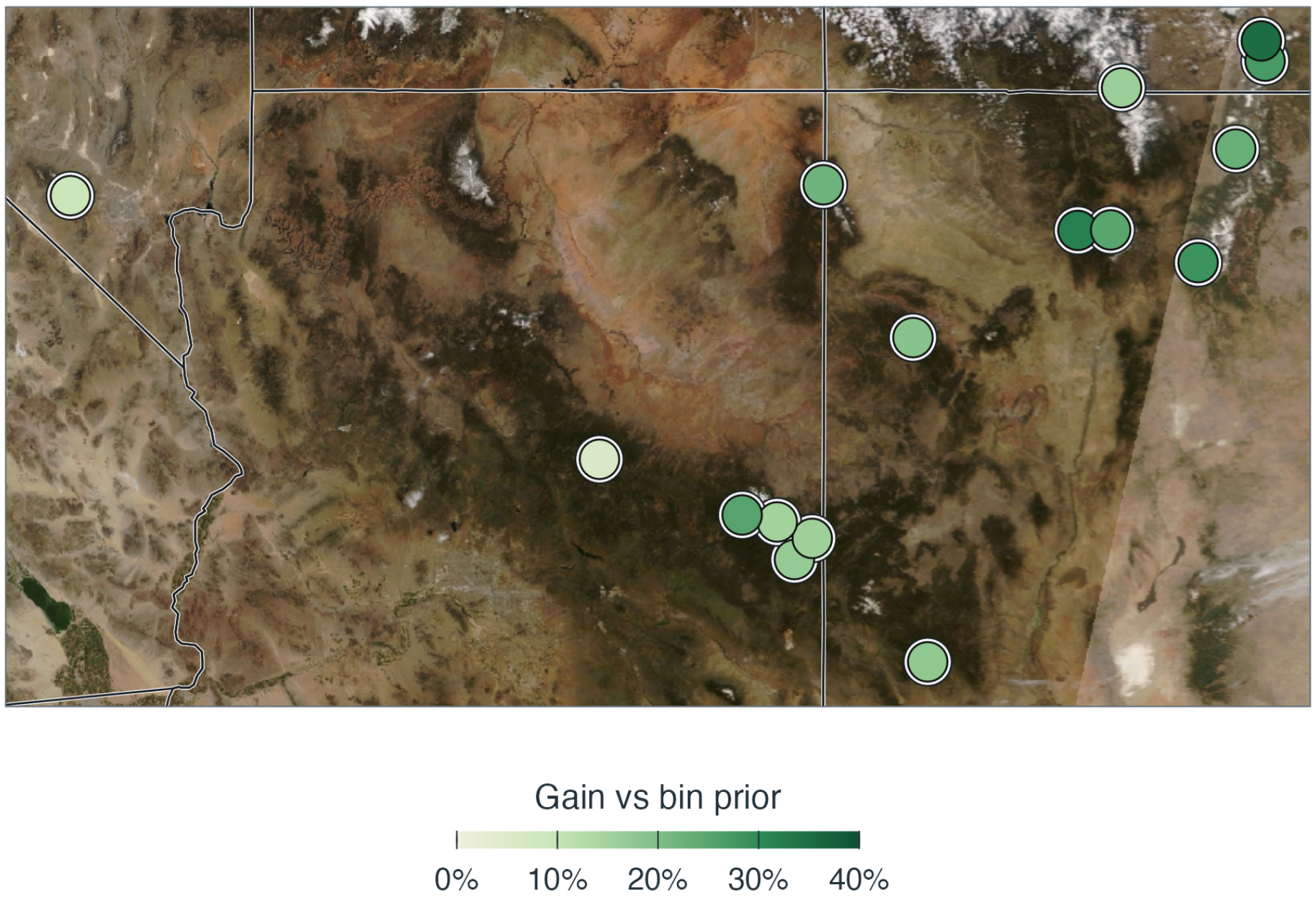

Figure 3. Held pillows on cloudless Southwest true colour, coloured by relative RMSE gain versus the seasonal change prior.

The map uses the same study box as Figure 2 with a cloudless MODIS true colour basemap and thin state outlines for orientation. Each of the 16 held pillows is a coloured point; the fill is relative RMSE gain of the proposed forecast for target date $t + 7$ over the seasonal change prior, pooled over held days at that pillow. The colour bar runs from low to high fractional gain. No pillow on the map entered fitting, and this is the only lon lat map in the paper that colours these sites by relative RMSE gain. Gain is positive at every held pillow, from 5.7 percent to 35.2 percent with a mean of 20.9 percent. Larger gains sit on pillows where the calendar window prior misses storm and melt timing that weather and neighbour state help resolve. Because every held site improves, the pooled 20.7 percent relative RMSE gain is not the product of a few mountain outliers.

### *4.2 Multi product map figures*

Figures 4 through 6 are three five by five map comparison figures that each show five forecast dates as columns. Each column pairs forecast date $t$ with target date $t + 7$: the column header is the forecast date month and year, the proposed row is the densified forecast map of SWE on target date $t + 7$, and the pillow row is pillow SWE on that same target date. The SNODAS and UA rows are daily analysis and reconstructed fields on the forecast date $t$; they are not multi day SWE forecasts, not target date product fields, and not the relative RMSE gain baselines. Placing the seven day forecast surface beside those analyses tests whether the densified map recovers comparable mountain versus desert structure without ingesting product SWE. True colour uses the clearest nearby acquisition within about ten days of the forecast date. Shared extents and a shared SWE colour scale make structure comparable without chopping one continuous field into place tiles. Figure 4 is the Southwest map comparison on the full study box.

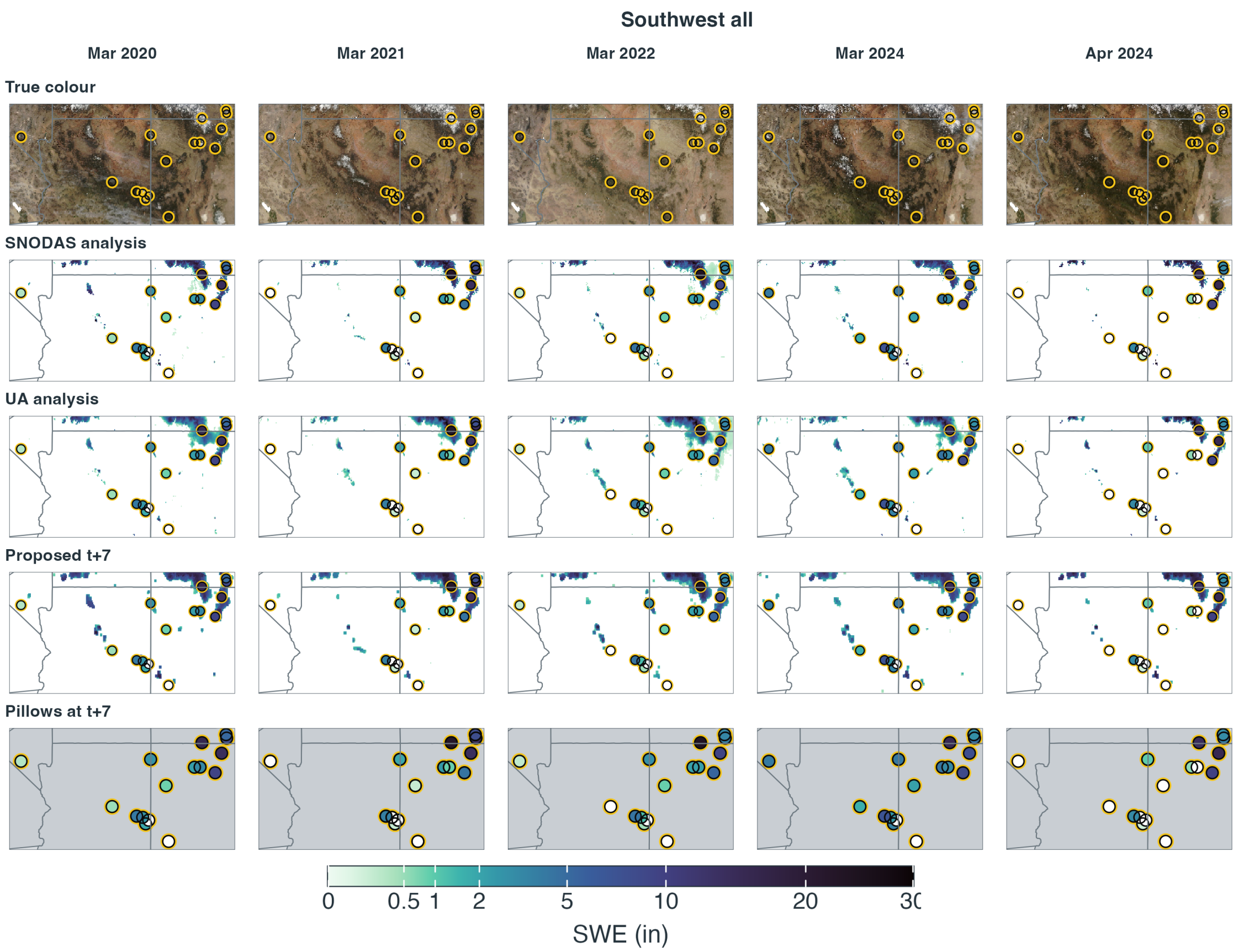

Figure 4. Southwest map comparison: forecast date columns with true colour, SNODAS and UA analyses, proposed $t + 7$ forecast map, and pillows on the target date $t + 7$.

Columns are forecast dates in March 2020, March 2021, March 2022, March 2024, and April 2024 inside the full Southwest box of Figure 2. Rows are true colour with pillow markers, SNODAS analysis, UA analysis, the proposed $t + 7$ forecast map, and pillows on the target date $t + 7$, all on one shared SWE colour scale. The March 2022 column is the clearest optical comparison case: bright snow on true colour marks the high ranges where the proposed $t + 7$ forecast and the SNODAS analysis both concentrate water equivalent while deserts stay near zero. That match is a geographic quality check of a seven day forecast against daily analyses, not a claim that SNODAS or UA themselves forecast $t + 7$.

Figure 5 zooms to the Sangre de Cristo window with the same forecast date versus target date row meanings and late winter forecast dates.

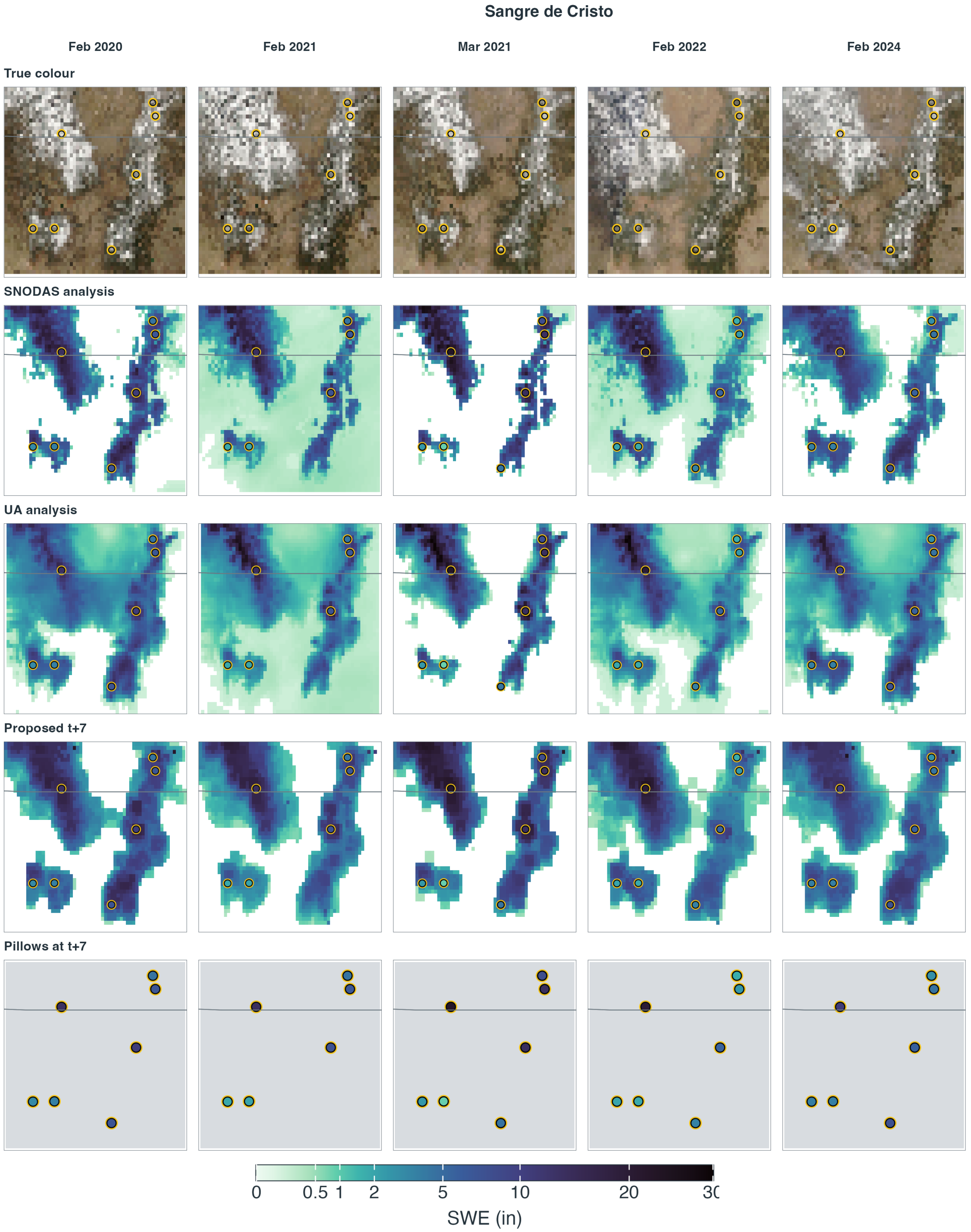

Figure 5. Sangre de Cristo map comparison with the same forecast date columns, proposed $t + 7$ map, and pillows on the target date $t + 7$ as Figure 4.

Columns share one fixed lon lat extent on the Sangre de Cristo window outlined in Figure 2. Row order matches Figure 4 so mountain cores can be compared without changing which calendar day each row represents. The proposed surface is densified with the terrain elev to SWE workflow and anti false snow steps from weather and remote sensing covariates, not from SNODAS plus a short residual field. Snow concentrates on the high ridge line while the lower floors stay near zero, which is the pattern a water manager would recognise against the SNODAS and UA analyses in the same frame.

Figure 6 applies the same layout to the Mogollon Rim and White Mountains window.

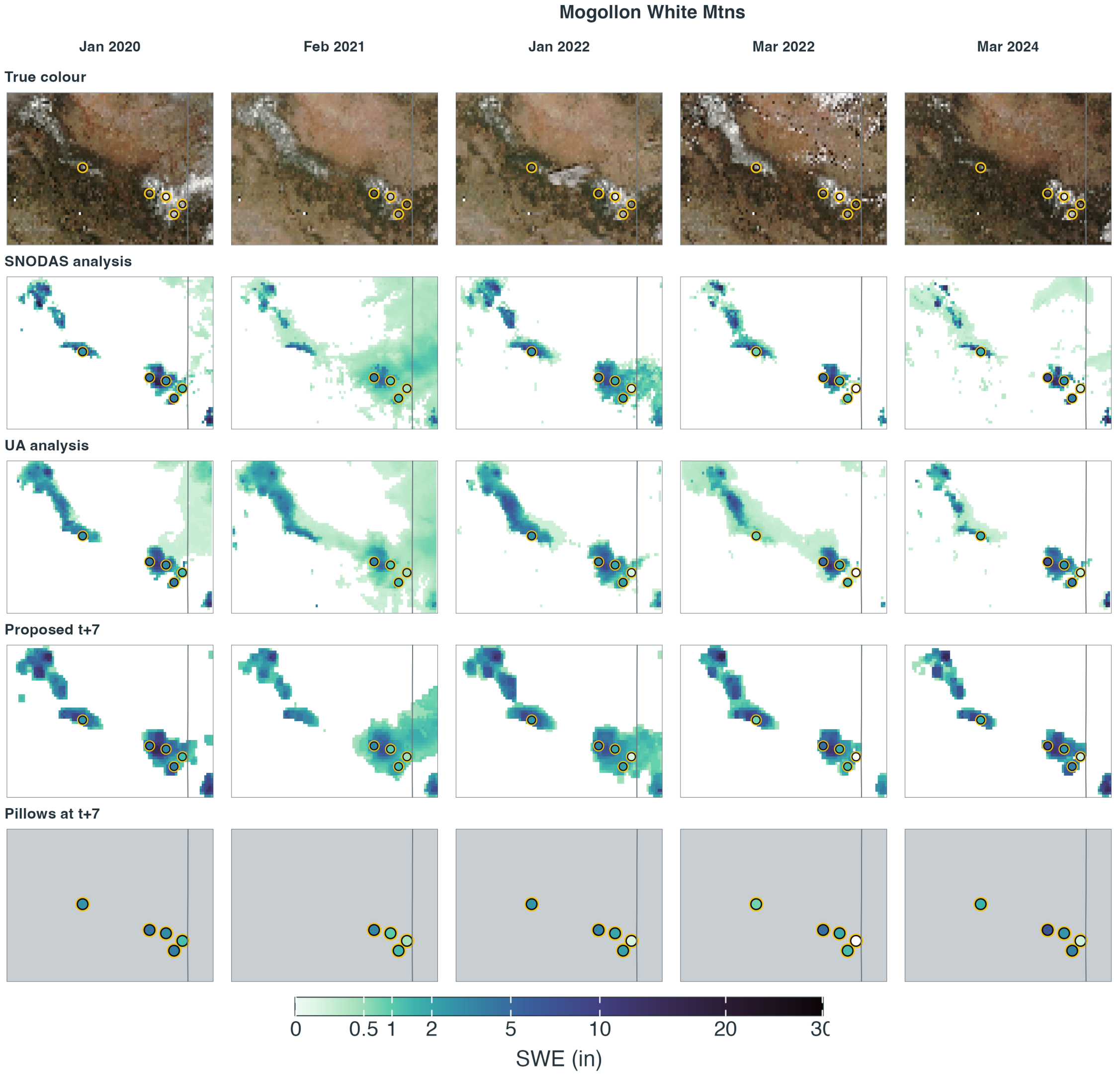

Figure 6. Mogollon Rim and White Mountains map comparison with the same forecast date versus target date rows as Figure 4.

Pillow markers on the true colour row locate the network inside this Arizona New Mexico window. Pillow colours on the bottom row are pillows on the target date $t + 7$ that the forecast must approach. The proposed $t + 7$ row is the densified forecast map for that same target date and again follows DEM ridge corridors rather than spraying MixGCN station blobs across the grid, so mid elev shelves stay off while ridge and slope snow remain visible beside the SNODAS and UA analysis rows.

In all three map comparison figures the SNODAS and UA analyses concentrate SWE on the high ranges while deserts stay near zero. The proposed seven day forecast surface follows that mountain versus desert contrast through DEM ridge corridor and slope densification rather than through MixGCN station blobs: snow grows along elev prominence slope spines, flat shelves stay off, sparse days refill lapse cooled ridges, and tip depth restore keeps high elev cores near SNODAS analysis tip means after morphology is fixed. Depth is written only where snow is on. Desert and foothill cells stay near zero, ridge and slope lines remain visible beside the analysis rows, salt and pepper freckles stay rare, and the held table still shows lower pillow RMSE for the proposed station forecasts than for the seasonal change prior. Producing a target date forecast map that approaches the geographic quality of those daily analyses, without using them as inputs, is the map product claim that accompanies the pillow relative RMSE gain.

Across those map dates the correlation of the proposed $t + 7$ forecast with the SNODAS analysis on the forecast date has mean about 0.79 and minimum about 0.67 on snowy cells, which is strong enough for shared mountain versus desert geography and still well below 0.95, so the mapped field is not a SNODAS wallpaper twin. Contour following elev to SWE depth on DEM ridge corridors uses an inside mask shallow to deep taper, scaled by relative distance to the pack edge, so pack edges and snowline bands stay light while ridge cores stay deep, rather than a hard jump from zero to deep pack or a soft buffer ring outside the mask (mean elev depth correlation on snow cells about 0.62). About 0.24 of snow cells fall in the light band below 2 in. Salt and pepper freckles are below 0.002 of snow cells, false snow on SNODAS bare cells is about 0.066 of domain cells, SNODAS snow recall is about 0.99, and the mean SWE ratio on the top tenth of snowy elev cells is about 1.19, so speckled noise and no snow overestimation stay controlled while tip cores remain deep, without SNODAS or UA inputs.

Figure 7 summarises the same three geographic windows on one forecast date each before the held pillow RMSE comparison begins.

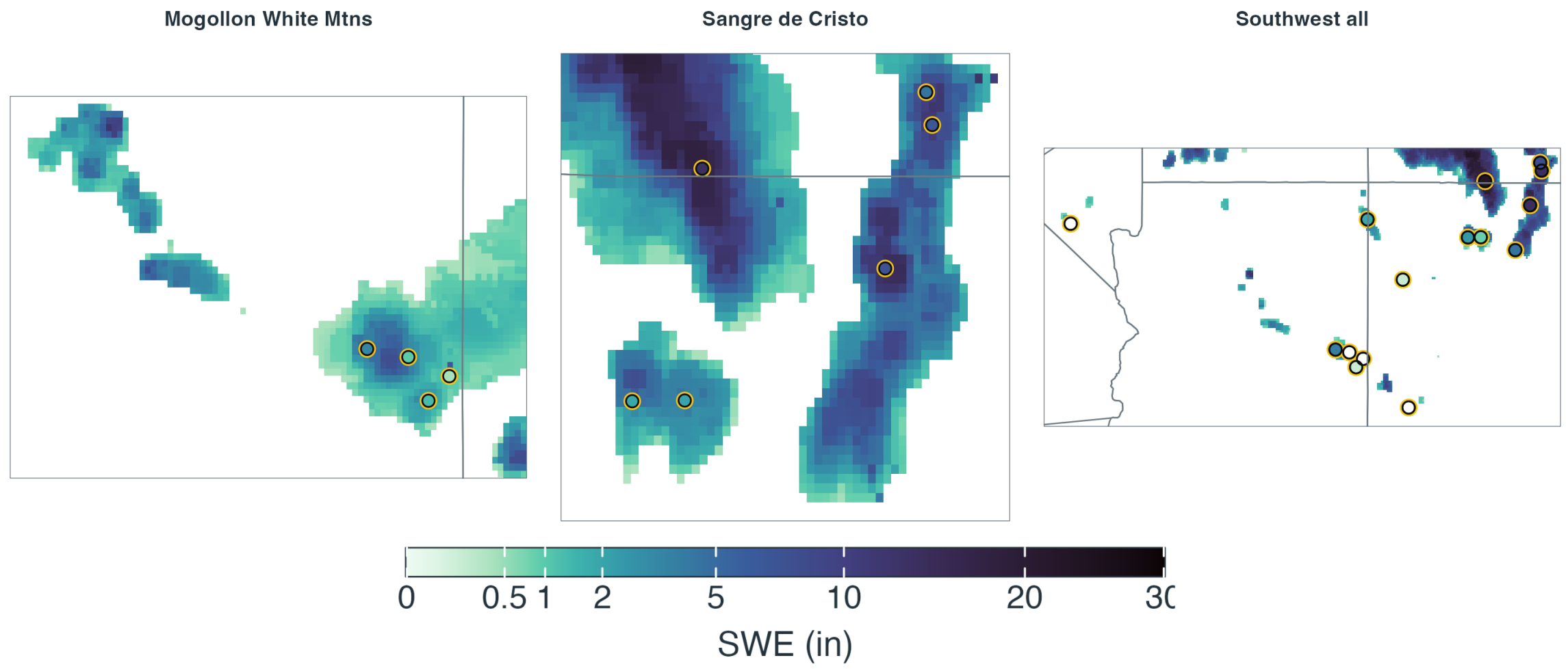

Figure 7. Single forecast date snapshots for the Southwest, Sangre de Cristo, and Mogollon windows with true colour, proposed $t + 7$ overlay, and pillows.

Three side by side maps therefore summarise the same three geographic windows used in Figures 4 through 6 on one forecast date each, with true colour under the proposed $t + 7$ forecast overlay and pillow markers. Densified mountain snow and bare desert floors remain aligned across zoom levels, which clears the geographic claim before the held pillow RMSE comparison begins.

*4.3 Classical forecast date baselines on one scale*

Geographic agreement alone does not prove forecast accuracy, so the next question is whether a classical forecast date prior already matches held pillows closely enough that a learned map would add little. Figure 8 ranks persistence, the seasonal change prior, anomaly persistence, and architecture ablations beside the proposed network on the same 11006 days.

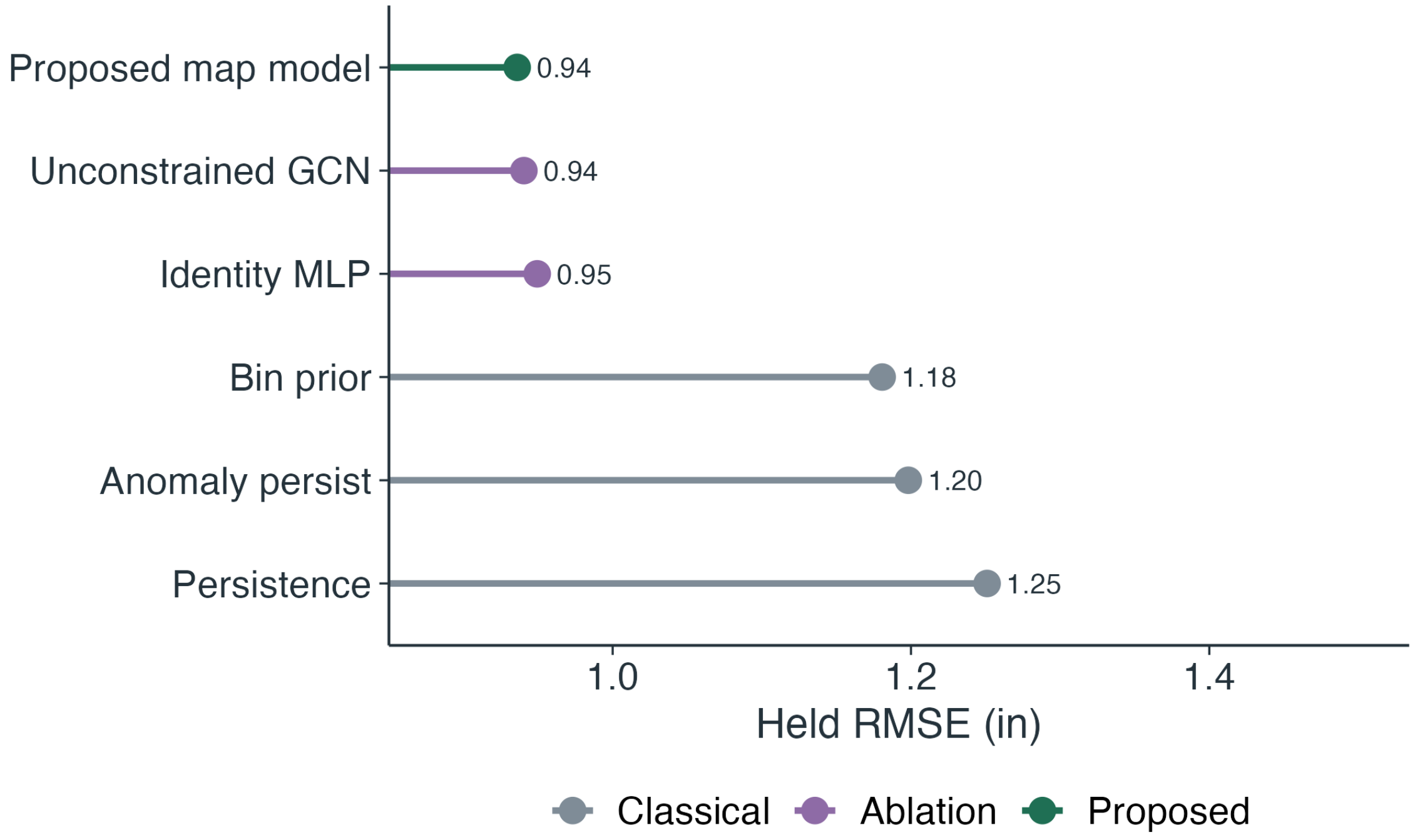

Figure 8. Held RMSE lollipop chart for classical baselines, ablations, and the proposed map model.

Each horizontal stem is a method evaluated on the same 11006 pillow days at 16 held pillows; the axis is RMSE in inches with values labelled to two decimals. Classical forecast date baselines (persistence, seasonal change prior, anomaly persistence, and related priors) sit above the architecture ablations and the proposed soft constraint MixGCN. The seasonal change prior has RMSE 1.181 inches and is the strongest classical baseline. Persistence is 1.251 inches. The unconstrained graph and identity network are 0.940 and 0.949 inches. The proposed network is 0.936 inches. Relative RMSE gain versus the seasonal change prior is 20.7 percent, which clears the 0.20 relative RMSE gain threshold. SNODAS and UA are not on this ladder because they are forbidden as inputs and are not the primary forecast baselines for this paper.

*4.4 Held years*

That pooled 20.7 percent relative RMSE gain could still be carried by one winter, so Figure 9 compares the proposed model with the seasonal change prior and persistence by held year.

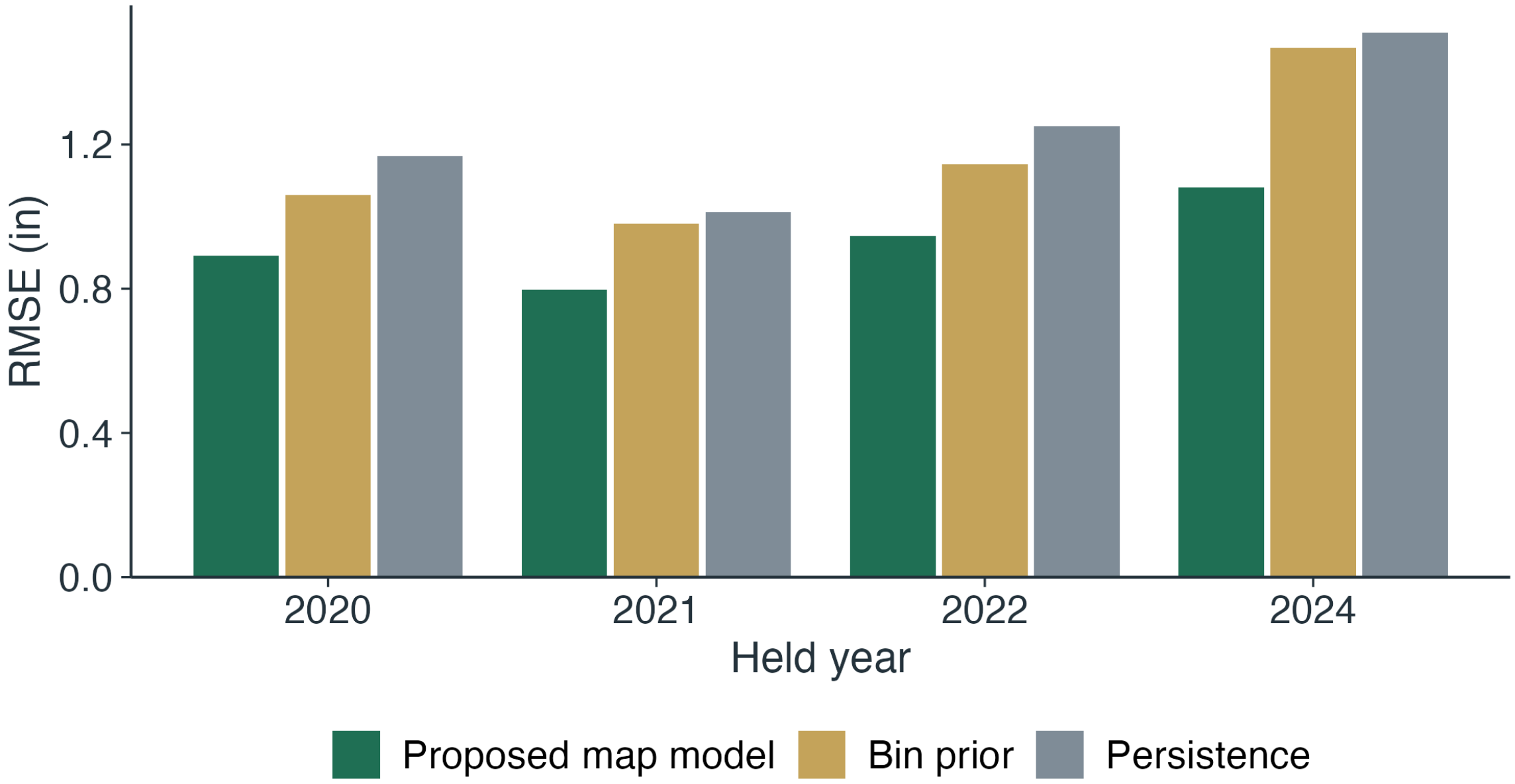

Figure 9. Held RMSE by year for the proposed model, the seasonal change prior, and persistence.

Grouped bars compare three methods on each of the four held winters (2020, 2021, 2022, and 2024) using the same held pillows as Figure 8. Proposed RMSE is 0.891, 0.797, 0.947, and 1.080 inches across those years. The seasonal change prior is 1.060, 0.981, 1.145, and 1.468 inches, so year level relative gains are 15.9, 18.7, 17.4, and 26.4 percent. Every held winter stays below both classical references, so the pooled 20.7 percent relative RMSE gain is not carried by a single winter.

*4.5 Observed versus forecast density*

Year level RMSE still leaves open how errors sit across the SWE range, so Figure 10 places observed target SWE against the proposed forecast and against persistence as hexagonal density on shared axes.

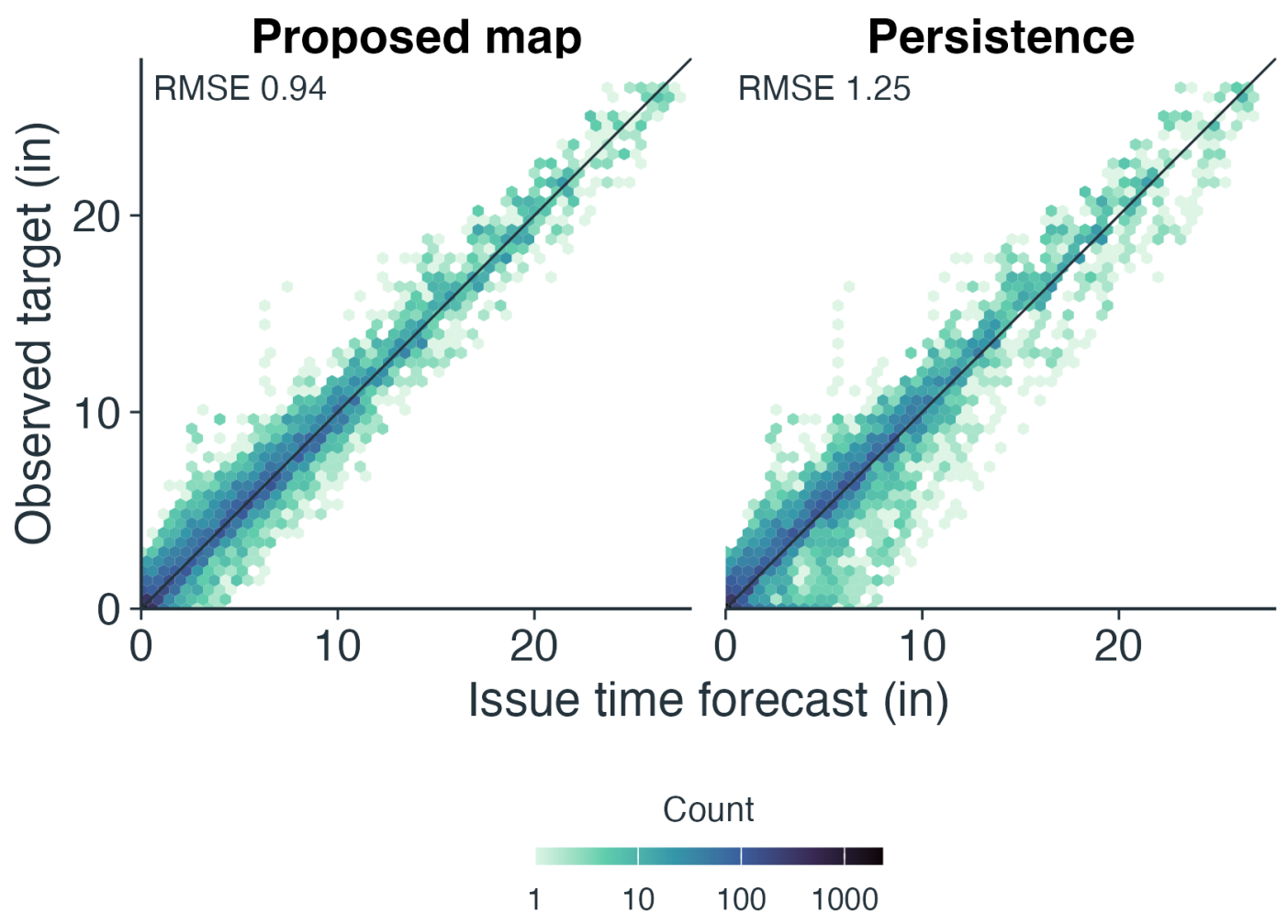

Figure 10. Hexagonal density of observed target SWE versus proposed and persistence forecasts.

Two side by side hexbin plots share axes from 0 to 28 inches with a one to one line; colour is count on a logarithmic scale. The left subplot is the proposed forecast (RMSE 0.94 inches) and the right subplot is persistence (RMSE 1.25 inches) on the same held pillow days. The proposed cloud is tighter about the one to one line at mid and high SWE than persistence, which holds forecast date SWE fixed and therefore lags storm and melt days. That pattern matches the RMSE gap on the lollipop chart and shows where absolute level agreement alone would overstate persistence on an autocorrelated target.

### *4.6 Initial SWE strata*

The hexbin already suggests mid and deep packs matter, so Figure 11 checks whether the gain is carried only by near zero packs using strata by initial SWE depth.

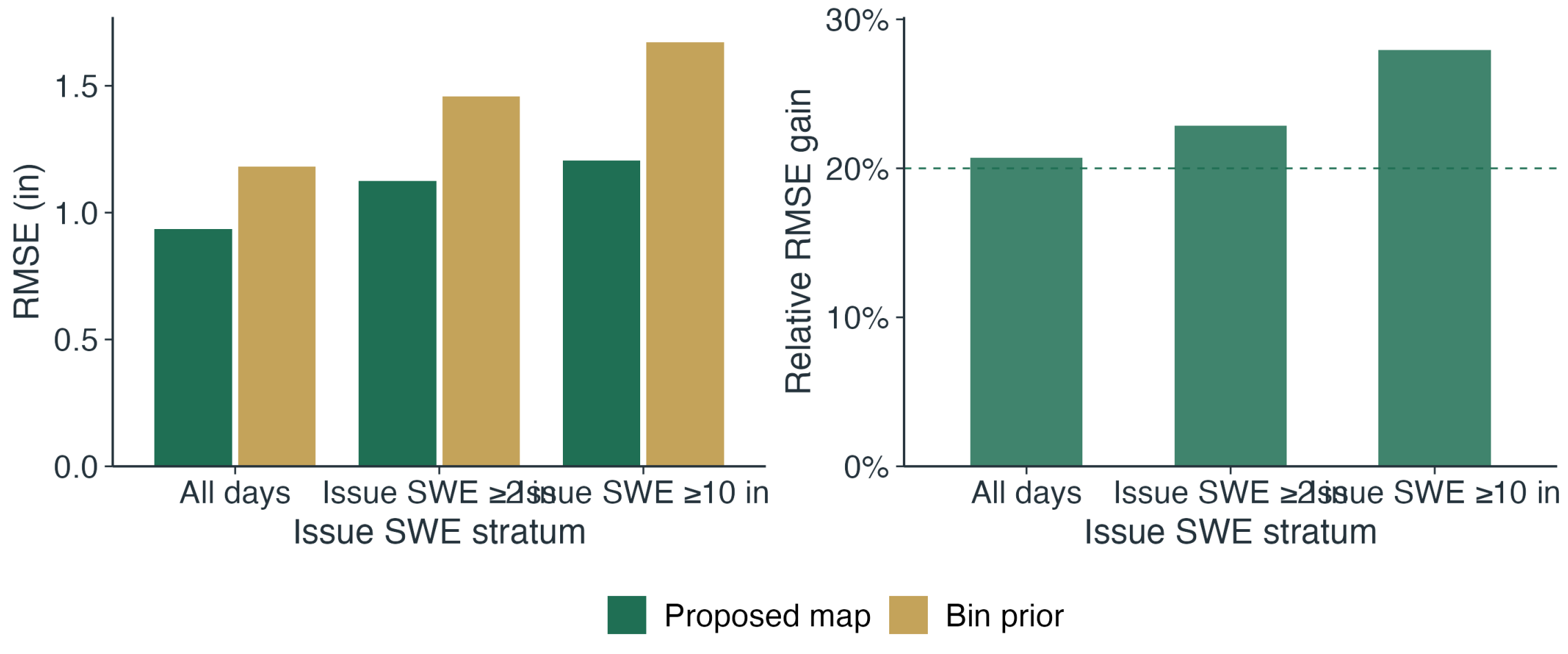

Figure 11. Held RMSE and relative RMSE gain versus the seasonal change prior by initial SWE stratum.

The left subplot is RMSE and the right subplot is relative RMSE gain for three strata: all days ($n = 11006$), initial SWE of 2 inches or more ($n = 6450$), and deep snow of 10 inches or more ($n = 1640$). The right subplot marks a 20 percent gain line. Relative RMSE gain versus the seasonal change prior is 20.7 percent on all days, 22.9 percent when initial SWE is at least 2 inches, and 27.9 percent on deep snow. The gain threshold is therefore not carried by zero snow rows; deeper packs, where water supply decisions matter most, show the larger fractional improvement.

*4.7 Signed bias*

Lower RMSE on deep snow could still hide a cold bias, so Figure 12 places mean signed error and underestimation rates for the proposed forecast beside persistence and the seasonal change prior.

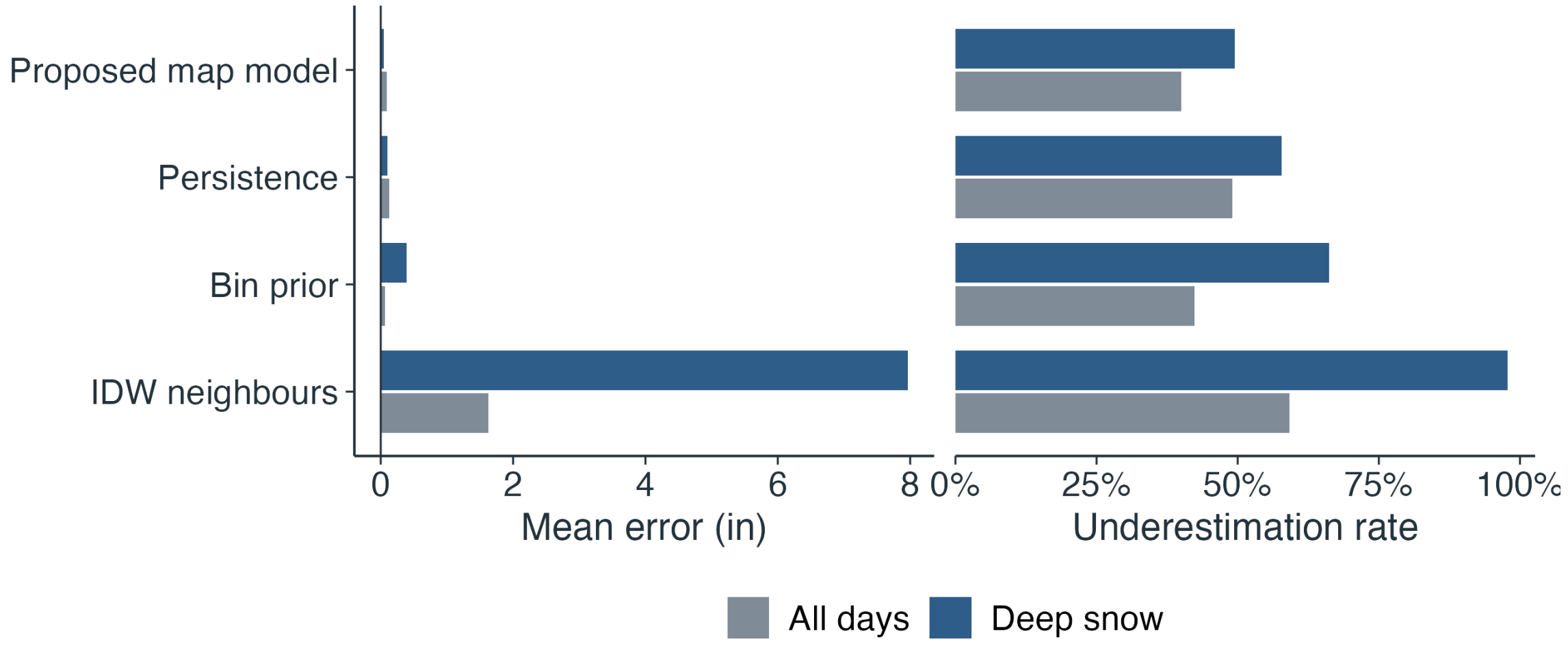

Figure 12. Mean forecast error and underestimation rate for the proposed model, persistence, and the seasonal change prior.

Two subplots compare the three methods on all held days and on deep snow days. The left subplot is mean signed error in inches; the right subplot is the fraction of days where the forecast is below the pillow. The proposed mean error is near zero overall. Persistence underestimates less often on deep snow than the seasonal change prior, which is consistent with mean reversion in the calendar window prior on accumulation days, while the soft constraint MixGCN remains closer to the pillow on average. The asymmetric underprediction weights in training therefore do not trade the RMSE win for a large cold bias on the held sample.

*4.8 Winter time series*

Pooled bias metrics still hide storm and melt timing, so Figure 13 shows daily series at three held pillows in January through April 2022. Each horizontal tick is a target date; the proposed curve at that day is the forecast produced on day $t$ for target date $t + 7$.

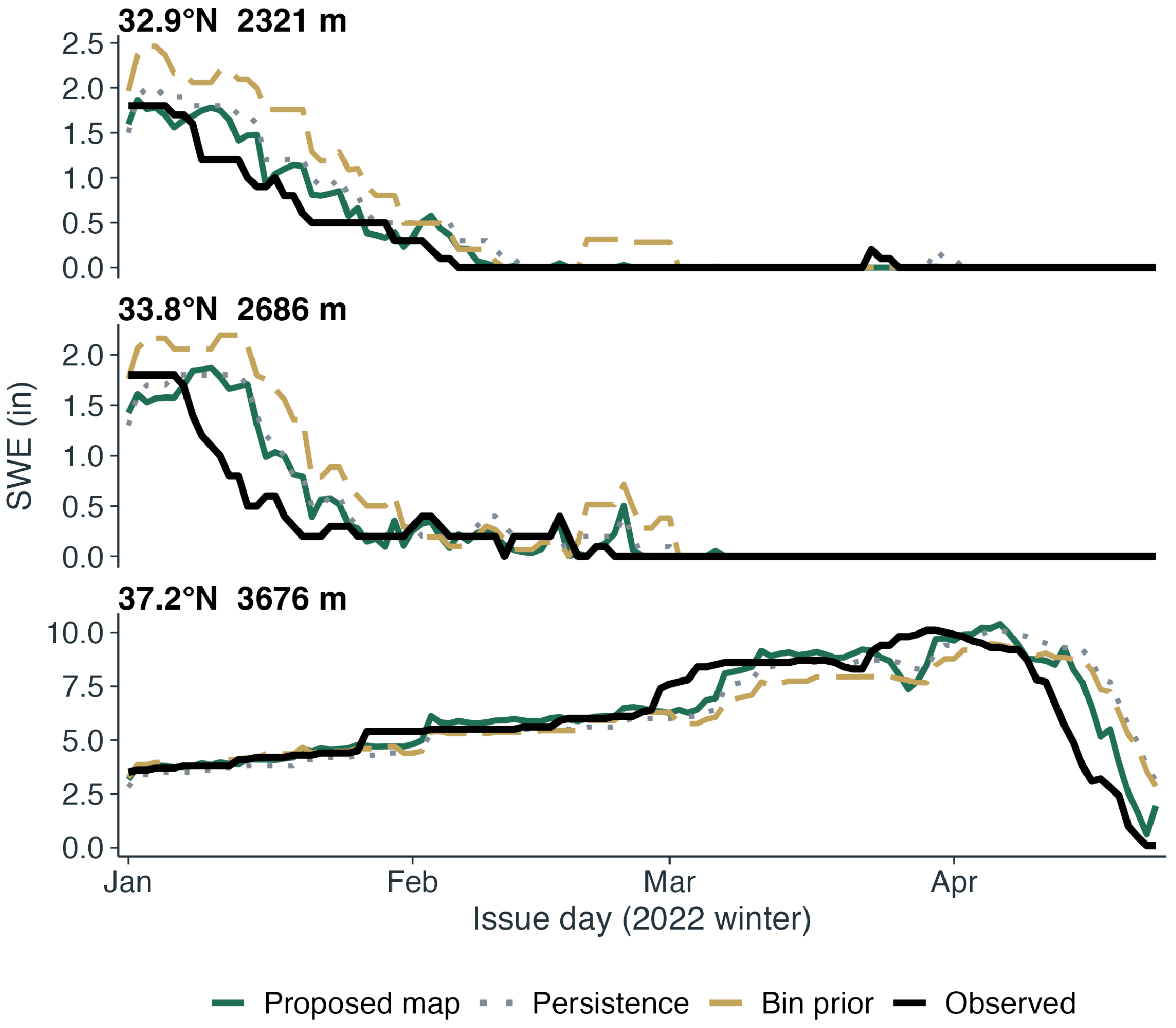

Figure 13. January to April 2022 target date time series at a low, mid, and high held pillow.

Three stacked subplots share a January through April 2022 window with sparse month ticks on the horizontal axis and SWE in inches on the vertical axis. Curves are the observed pillow on the target date, the proposed forecast for that target date, persistence (forecast date initial SWE held fixed to the target date), and the seasonal change prior for the same target date. The low, mid, and high elev pillows show the same qualitative pattern: the proposed forecast tracks storm rises and melt falls more closely than persistence, which cannot change between forecast date and target date, while the seasonal change prior often under or over shoots when the calendar window mean change does not match that winter. The series therefore illustrate, at daily resolution, the RMSE and hexbin advantages reported in the pooled held sample.

*4.9 Distance to the fitting network*

The final held check asks whether those timing gains survive away from the dense fitting neighbourhood. Held pillows are stratified by distance to the nearest fit pillow inside the study domain of Figure 2.

Relative RMSE gain versus the seasonal change prior is 20.8 percent within 30 km of a fit pillow ($n = 9612$), 17.7 percent between 30 and 80 km ($n = 696$), and 18.9 percent beyond 80 km ($n = 698$). The farthest distance band still clears a large positive gain, which is the check that a model trained only on dense pillows would be expected to weaken and the quantitative check on spatial generalisation within this southwestern experiment.

## 5 Discussion

Taken together, the map comparison figures and the held pillow scores answer the same question posed in the Introduction: a soft constrained MixGCN can produce a densified forecast map of SWE on target date $t + 7$ from weather and remote sensing inputs without ingesting SNODAS or UA, and those forecasts reduce held pillow RMSE by 20.7 percent relative to the best classical forecast date baseline on a spatially withheld sample inside the southwestern study domain of Figure 2. That relative RMSE gain comparison is the pillow forecast claim: both sides may use only information available on the forecast date, scores are on target date $t + 7$, and the proposed side never sees operational SWE analyses as features. The geographic map claim is separate and complementary: SNODAS and UA are daily analysis and reconstructed fields, not seven day forecasts, and the densified target date surface is judged against those analyses for geographic quality rather than as if the products themselves were forecasting competitors.

The reason the maps are not a pillow twin is the pairing of those fitting steps for cells without ground stations with terrain densification. Hiding private pillow SWE history and blanking nearby fit pillow groups force the forward pass to resemble cells with no private archive during fitting. Scoring temporary locations without ground stations and raising loss weight at isolated fit pillows push the soft constraint toward weather driven physics away from dense SNOTEL clusters. The map surface then uses DEM elev to SWE morphology, prominence anti false snow, lapse corrected sparse ridge rescue, and tip depth restore rather than pasting SNODAS plus a short residual field or spraying MixGCN pillow predictions across the grid. Held pillows beyond 80 km of the nearest fit pillow still show an 18.9 percent relative RMSE gain versus the seasonal change prior, which is the measurable distance check that the map claim requires.

SNODAS and UA remain useful geographic references in those map figures precisely because they are daily analyses, not forward forecasts. Treating their target date fields as competitors to pillow forecasts is a different experiment and is not the relative RMSE gain comparison here; feeding those fields into the network would convert the model into a product corrector. The densified seven day forecast surface shares mountain versus desert geography with the SNODAS analysis on the forecast date at mean snowy cell correlation about 0.79 while remaining below the 0.95 wallpaper threshold, so map product distinctness remains visible without ingesting product SWE. Contour following elev to SWE depth on DEM ridge corridors, with an inside mask shallow to deep taper scaled by relative pack edge distance and without soft rim buffers outside the snow mask, keeps light edge pack and deep ridge cores visible beside SNODAS and UA. Salt and pepper freckles stay below 0.002, false snow on SNODAS bare cells is about 0.066 of domain cells, SNODAS snow recall is about 0.99, and the top tenth elev SWE ratio is about 1.19, which jointly shows that a weather and remote sensing forecast map can approach analysis grade mountain structure under forecast date information rules.

The soft constraint weight selected on validation winters is 0.15. Matched ablations show the soft constraint MixGCN beats both the identity network and the $\lambda = 0$ graph on the same held sample, so the gain is not explained by ensemble averaging alone. Weather and neighbour state supply the information that replaces product analyses on the forecast date.

Limits that change interpretation include the single draw of 16 held pillows, the southwestern domain, the fixed target date at $t + 7$, and the fact that temporary training locations without ground stations still borrow weather from fit pillows rather than from a fully independent meteorological stack on every cell. Soft constraint weight selection uses validation winters only, but architecture search was not blind to held behaviour. Three assumptions weakened by residual errors are that calendar window priors already capture seven day storm and melt timing, that weather and snow index fully replace assimilation texture far from pillows, and that a southwestern held pillow draw generalises to the full western network.

## 6 Conclusions

A soft constrained mix adjacency graph network trained on SNOTEL pillows from one $F = 52$ feature vector per forecast date, built with up to 21 days of pillow SWE and weather history ending on that day plus neighbour SWE, terrain, snow index, and forecast date priors, produces a densified regional SWE map for target date $t + 7$ from each forecast date $t$, and those forecasts reduce held pillow RMSE by 20.7 percent relative to the best classical forecast date baseline, the seasonal change prior, on 11006 held pillow days. Fitting randomly hides private pillow SWE history, blanks nearby fit pillow groups, scores temporary locations without ground stations, and raises loss weight at isolated fit pillows so the checkpoint serves map cells with no private archive rather than memorising dense SNOTEL sites alone, and the map surface is written by a separate terrain densification workflow rather than by a pillow only forward pass. SNODAS and UA are daily analysis and reconstructed fields, not seven day SWE forecasts, and they never enter the model. The gain is positive at every held pillow and in every held year, remains positive beyond 80 km from the fitting network, and is larger on deep snow. Multi product grids place the densified target date forecast beside true colour, SNODAS analysis, UA analysis, and target date pillow observations so geographic quality of the seven day forecast can be judged against those analyses. The proposed maps share mountain versus desert geography with the SNODAS analysis at mean snowy cell correlation about 0.79, keep ridge and slope lines visible through contour following DEM corridor densification with an inside mask shallow to deep taper, hold salt and pepper freckles below 0.002, hold false snow near 0.066, retain about 0.99 SNODAS snow recall, place about 0.24 of snow cells in the light band below 2 in, and keep high elev tip depth near a ratio of 1.19, which shows that a forward forecast map can approach analysis grade SWE geography without using SNODAS or UA as inputs.

*Future work*

Future work should expand meteorology at temporary training locations without ground stations beyond inverse distance weighted fit pillow banks, widen the domain beyond the Southwest, and add independent snow course or airborne lidar collocations where licensed data exist.

## Author contributions

Ziheng Sun: Conceptualization, Data curation, Formal analysis, Funding acquisition, Investigation, Methodology, Software, Validation, Visualization, Writing original draft, Writing review and editing. Hui Fang: Data curation, Software, Validation, Writing review and editing.

## Data availability

Supporting data and held predictions will be shared with the editor and reviewers upon request during peer review. The same materials will be published in a public archive after peer review.

**Funding**

This work was partially supported by NASA ACCESS (80NSSC21M0028), NSF RISE (2425687), NOAA (NA22OAR4590512, NA23OAR4590388, NA23OAR4590387), and NSF CyberTraining (2117834).

**Acknowledgements**

The authors thank colleagues who discussed map product evaluation design. Language editing assistance is disclosed in the Declaration of AI section below.

**Declaration of AI and AI-assisted technologies in the writing process**

During the preparation of this work the authors used generative AI assisted language tools in order to improve readability and language. After using this tool/service, the authors reviewed and edited the content as needed and take full responsibility for the content of the publication.